\documentclass[sn-mathphys-num]{sn-jnl}

\usepackage{graphicx}%
\usepackage{multirow}%
\usepackage{amsmath,amssymb,amsfonts}%
\usepackage{amsthm}%
\usepackage{mathrsfs}%
\usepackage[title]{appendix}%
\usepackage{xcolor}%
\usepackage{textcomp}%
\usepackage{manyfoot}%
\usepackage{booktabs}%
\usepackage{algorithm}%
\usepackage{algorithmicx}%
\usepackage{algpseudocode}%
\usepackage{listings}%
\usepackage{bm}%

\theoremstyle{thmstyleone}%
\theoremstyle{thmstyletwo}%

\theoremstyle{thmstylethree}%

\begin{document}

\title[Article Title]{Influence of Magnetic Field on Surface Andreev Bound States in Superfluid $^3$He-B Studied by Mobility of Electron Bubble}


\author*[1]{\fnm{Hiroki} \sur{Ikegami}}\email{hikegami@iphy.ac.cn}
\equalcont{These authors contributed equally to this work.}

\author[2]{\fnm{Kimitoshi} \sur{Kono}}\email{kkono@nycu.edu.tw}

\author*[3]{\fnm{Yasumasa} \sur{Tsutsumi}}\email{y.tsutsumi@kwansei.ac.jp}
\equalcont{These authors contributed equally to this work.}

\affil[1]{\orgdiv{Beijing National Laboratory for Condensed Matter Physics, Institute of Physics}, \orgname{Chinese Academy of Sciences}, \orgaddress{\city{Beijing}, \postcode{100190}, \country{China}}}

\affil[2]{\orgdiv{Department of Electrophysics and Center for Emergent Functional Matter Science}, \orgname{National Yang Ming Chiao Tung University}, \city{Hsinchu}, \postcode{300093}, \country{Taiwan}}

\affil[3]{\orgdiv{Department of Physics}, \orgname{ Kwansei Gakuin University}, \orgaddress{\street{Sanda}, \city{Hyogo}, \postcode{669-1337}, \country{Japan}}}


\abstract{
The B phase of superfluid $^\textrm 3$He ($^\textrm 3$He-B) is topologically nontrivial and the surface Andreev bound states formed on a surface are conceived as Majorana fermions.
In a magnetic field, the surface Andreev bound states acquire a Zeeman gap.
How the Zeeman gap opens when a magnetic field is applied is intimately related to how the topological properties are lost.
In this article, we study the mobility of an electron bubble trapped under a free surface of $^\textrm 3$He-B in a magnetic field of 0.25~T to examine the influence of the magnetic field on the surface Andreev bound states.
We observe experimentally and theoretically that, with decreasing temperature, the mobility at 0.25~T increases steeper than that in zero magnetic field when the thermal energy is comparable to the Zeeman energy, while the mobility is slightly smaller than that in zero magnetic field at higher temperatures.  
These features are understood by the opening of the Zeeman gap, the resulting change in the density of states within the bulk superfluid gap, and the distortion of the bulk superfluid gap by the magnetic field. 
}




\maketitle

\section{Introduction}\label{sec1}
The concept of topology has opened a new paradigm in condensed matter physics.
Remarkably, Majorana fermions are predicted in topological superfluids/superconductors~\cite{Read2000,Qi2009,Qi2011,Alicea2012,Beenakker2013,Sato2016,Sato2017}, and their exotic properties have attracted fundamental and technological interest.
Superfluid $^3$He~\cite{Vollhardt90,Dobbs00,Leggett_book} with a well established order parameter provides an ideal platform to unambiguously examine fundamental properties relevant to topology~\cite{Volovik2003,Mizushima2015,Mizushima2016}.
The B phase of superfluid $^3$He ($^3$He-B) is a topological superfluid protected by time-reversal symmetry and should have the surface Andreev bound states which are conceived as Majorana fermions~\cite{Schnyder2008,Chung2009,Qi2009,Volovik2009_1}.
So far, the presence of the surface Andreev bound states has been demonstrated by several different experiments, such as measurements of an acoustic impedance~\cite{Aoki2005,Okuda2012,Murakawa2009,Murakawa2011}, heat capacity~\cite{Choi2006}, and transport of electron bubbles~\cite{Ikegami2013-2,Tsutsumi2017,Ikegami2019}.

In a high magnetic field, this symmetry is broken and $^3$He-B becomes non-topological.
The surface Andreev bound states have a Zeeman gap and they are no longer Majorana fermions.
How the Majorana fermions become non-Majorana fermions is intimately related to how the topological properties are lost (see Sect. 2 for more details). 
In this article, we study the mobility of the electron bubble trapped under a free surface of $^3$He-B in a magnetic field of 0.25~T experimentally and theoretically to examine the influence of the magnetic field.

Our platform for studying the surface Andreev bound states is a free surface of superfluid $^3$He.
It is known that the energy spectrum of the surface Andreev bound states is sensitive to details of the surface condition for the reflection of quasiparticles~\cite{Zhang1988,Nagato1998,Nagai2008,Murakawa2011}.
At a specular surface, where quasiparticles are specularly reflected, the surface Andreev bound states have a well-defined linear dispersion relation called the Majorana corn.
As the surface becomes more diffusive, that is, more quasiparticles are randomly reflected, the dispersion relation becomes more smeared, and finally it becomes very complicated at the diffusive limit. 
Therefore, studies at a specular surface are desirable to obtain a clear understanding.
The free surface is a specular surface, which has been demonstrated by the mobility of a Wigner crystal formed on a free surface~\cite{Shirahama1997,Ikegami2006}.

Electron bubbles trapped under the free surface provide a unique probe for studying the properties of the surface Andreev bound states formed at the free surface. 
When electrons are placed in liquid He, the electrons form the state of electron bubble, where the electron is self-trapped in a spherical cavity with a radius
$R $ of about 2~nm~\cite{Fetter76}.
As demonstrated, the experimental mobility $\mu$ measured under the surface of $^3$He-B at zero field~\cite{Ikegami2013-2,Ikegami2019} quantitatively agreed with the theory which took into account the elastic scattering of the surface Andreev bound states~\cite{Tsutsumi2017}.
This agreement indicates that the electron bubble is an excellent probe to elucidate the properties of the surface Andreev bound states quantitatively.
In this article, we report on experimental and theoretical studies of the mobility of the electron bubble trapped under a free surface of $^3$He-B in a magnetic field $H$ of 0.25~T.
At this field, the Zeeman gap $\hbar \gamma H /2$ is 0.19~mK, which is comparable to the lowest temperature of our experiment (0.25~mK) ($\hbar$ is Planck's constant divided by $2\pi$
and $\gamma$ is the gyromagnetic ratio of the $^3$He nucleus).

\begin{figure}[th]
	\begin{center}
		\includegraphics[width=1\linewidth,keepaspectratio]{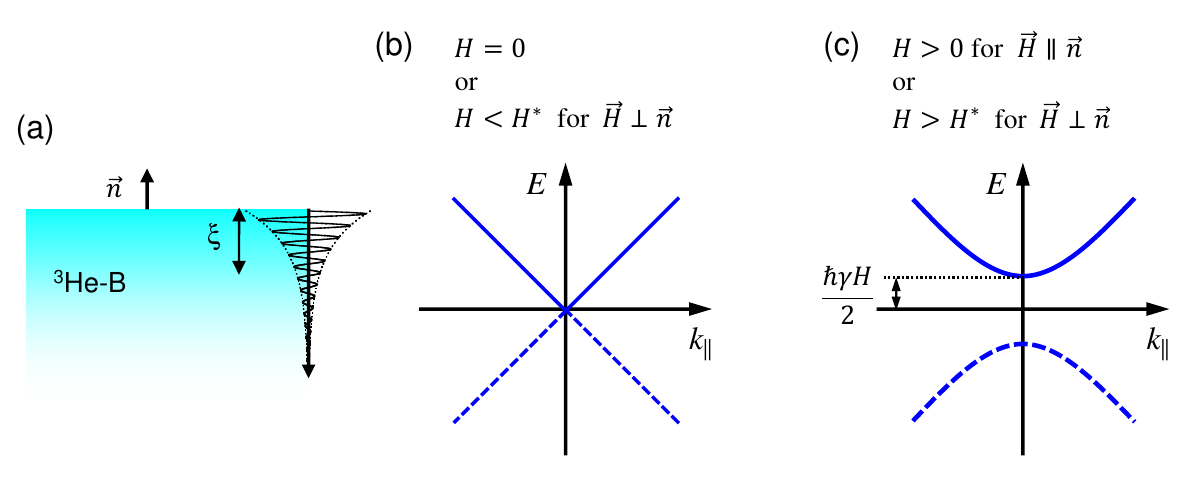}
	\end{center}
	\caption{(a) Surface Andreev bound states formed at a free surface. $\vec n$ denotes the surface normal and $\xi$ is the coherence length. (b) The dispersion relation of surface Andreev bound states. It is gapless at $H=0$ or $H < H^*$ for $\vec H \perp \vec n$. (Here $H=|\vec H|$) (c) Surface Andreev bound states have a Zeeman gap at $H>0$ for $\vec H \parallel \vec n$ or $H > H^*$ for $\vec H \perp \vec n$. }
	\label{Majorana_cone}
\end{figure}

\section{Surface Andreev bound states in magnetic field}

Surface Andreev bound states are formed at a free surface of bulk $^3$He-B within the coherence length (Fig.~\ref{Majorana_cone}a).
At zero magnetic field, they have a gapless linear dispersion (Fig.~\ref{Majorana_cone}b) and satisfy the Majorana condition~\cite{Chung2009,Schnyder2008}. 
When a magnetic field $H$ is applied perpendicular to the surface, the surface Andreev bound states acquire a Zeeman gap (Fig.~\ref{Majorana_cone}c). 
Their energy is given by $E_\mathrm{k_{\parallel}}= \sqrt{\Delta_\mathrm{B} (k_{\parallel}/k_\mathrm{F})^2+(\hbar \gamma H/2)^2}$ , and the surface Andreev bound states are no longer Majorana fermions.
($\Delta_\mathrm{B}$ is the superfluid gap of bulk $^3$He-B, $k_{\parallel}$ is the wavenumber of the surface Andreev bound states parallel to the surface, and $k_\mathrm{F}$ is the Fermi wavenumber.)
We note that the Zeeman gap opens even at a tiny field when it is applied perpendicular to the surface.
This is in sharp contrast to the case of $H$ applied parallel to the surface, where the surface Andreev bound states remain gapless until the magnetic field exceeds the critical field $H^*$($\sim$~3~mT) because of the protection associated with a hidden discrete symmetry (chiral symmetry)~\cite{Mizushima2012-1,Mizushima2012-2}.
This anisotropic response to the magnetic field is called the Ising spin~\cite{Chung2009}, which is one of the exotic features of the Majorana fermions. 
At $H>H^*$, this symmetry is broken, and $^3$He-B becomes non-topological with a non-zero Zeeman gap.
At $H \gg H^*$, the surface Andreev bound states have the Zeeman gap for any orientation of the magnetic field (Fig.~\ref{Majorana_cone}c).

The magnetic field in this study is 0.25~T applied perpendicular to the surface, which is much larger than $H^*$.
Therefore, the surface Andreev bound states are expected to have the Zeeman gap.
This study aims to elucidate the magnetic properties of the surface Andreev bound states at high fields.
In addition, this study could be a first step towards the demonstration of the Ising spin at low fields ($H < H^*$).

\section{Experiment: mobility of electron bubble in magnetic field}

The experiments were performed with the same setup as described in Ref.~\cite{Ikegami2013-2,Ikegami2013-1}.
We measured mobility using the Sommer-Tanner method~\cite{Sommer71} with a rectangular geometry.
An ac voltage $V_\mathrm{in}$ was applied to the input electrode, which induced motion of electron bubbles parallel to the surface.
The motion of electron bubbles induced the current on the output electrode $I_\mathrm{out}$, which was recorded to obtain the mobility.
The frequency $f$ of $V_\mathrm{in}$ was chosen in the range of 0.06$-$12~Hz depending on the magnitude of the mobility.
The mobility was deduced by fitting of $I_\mathrm{out}$ to the formula derived from the transmission-line analysis~\cite{Mehrotra1987}.
The electron bubbles were generated by field emission from carbon nanotubes \cite{Kawasaki2005}.
The depth of the trapped electron bubbles reported in this article was 35~nm from the surface.
The magnetic field of 0.25~T was applied normal to the surface.
Temperature of liquid $^3$He below 500~$\mu$K was determined directly from the density of thermally excited quasiparticles in bulk $^3$He-B by using a vibrating wire and a tuning fork located in the low-field $^3$He-B region.
Above 500~$\mu$K, a platinum NMR thermometer mounted on a nuclear stage was employed. 
Data were acquired by slowly sweeping temperature at a rate of less than 30~$\mu$K/h.
For further details, see Ref.~\cite{Ikegami2013-2}.

\begin{figure}[th]
	\begin{center}
		\includegraphics[width=0.8\linewidth,keepaspectratio]{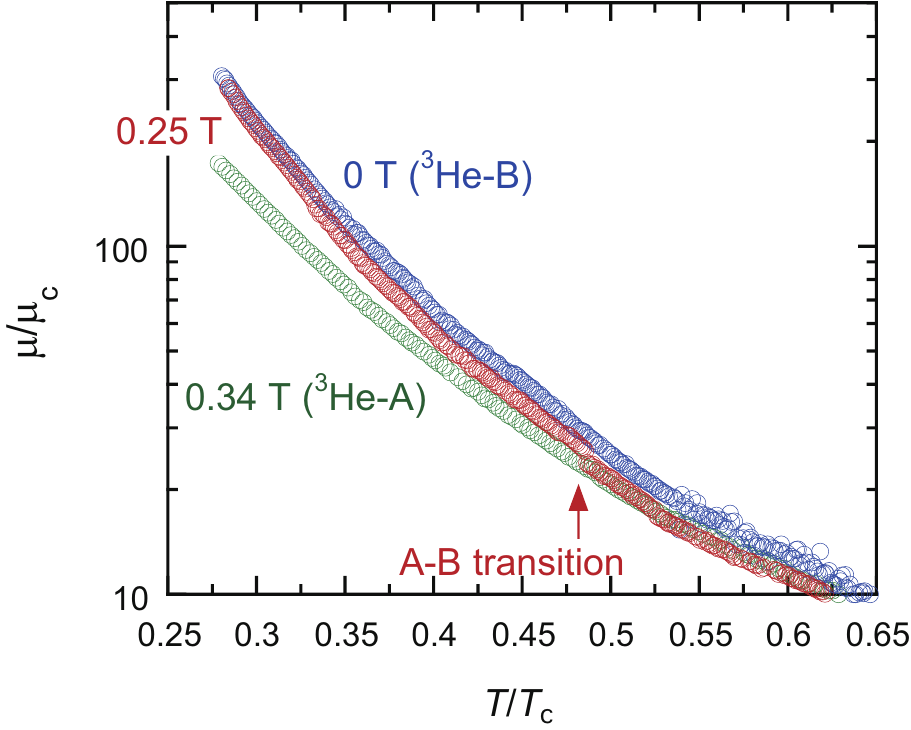}
	\end{center}
	\caption{Mobility of electron bubbles trapped under a free surface. The magnetic field is applied normal to the surface. At 0.25~T, $^3$He-B is realized at $T<$~0.48$T_{\rm c}$, while  $^3$He-A is formed at $T>$~0.48$T_{\rm c}$. For comparison, the mobility for $^3$He-B at 0~T and $^3$He-A taken at 0.34~T are shown~\cite{Ikegami2013-2}. $\mu_\mathrm{c}$($=$ 1.7$\times$10$^{-6}$~m$^2$/Vs) is the mobility at the superfluid transition temperature $T_{\rm c}(=$ 0.93~mK$)$. The depth of the electron bubbles is 35~nm. The data at 0.25~T are obtained on cooling, and the observed A$-$B transition temperature is lower than the thermodynamic transition temperature ($\sim$0.65$T_{\rm c}$)~\cite{Kyynarainen1990} due to supercooling.
	}
	\label{mu-T}
\end{figure}

The mobility of electron bubbles measured at 0.25~T is shown in Fig.~\ref{mu-T} as a function of temperature $T$.
In the figure, $\mu$ measured for $^3$He-B at 0~T and $^3$He-A at 0.34~T are also exhibited for comparison~\cite{Ikegami2013-2}.
At 0.25~T, the system is in $^3$He-A at $T>$~0.48$T_{\rm c}$, and $\mu$ shows the same temperature dependence as that for $^3$He-A measured at 0.34~T ($T_{\rm c}$ is the transition temperature of bulk superfluid $^3$He).
At $T=$~0.48$T_{\rm c}$, the transition from $^3$He-A to $^3$He-B occurs upon cooling, and $\mu$ shows a small jump.  
With decreasing temperature further, $\mu$ at 0.25~T approaches the mobility at 0~T at temperatures below $\sim$~0.35$T_{\rm c}$.

The mobility is inversely proportional to the drag force acting on the moving electron bubble.
Near the surface, the drag force has two contributions: scattering with thermally-excited bulk quasiparticles and that with surface Andreev bound states~\cite{Tsutsumi2017}.
The drag force caused by the scattering with bulk quasiparticles decreases quickly with decreasing temperature due to the exponential reduction of the density of bulk quasiparticles. 
On the other hand, the decrease of the drag force caused by the surface Andreev bound states upon cooling is slower because they are formed within $\Delta_\mathrm{B}$.  
In the case of zero magnetic field, the drag force arising from surface Andreev bound states is about 60\% of the total drag force at 0.27$T_{\rm c}$~\cite{Tsutsumi2017,Ikegami2019}.
Thus, the mobility at temperatures around 0.27$T_{\rm c}$ is largely determined by the surface Andreev bound states. 
At 0.25~T, the surface Andreev bound states have a Zeeman energy $\hbar  \gamma H /2$ of $\sim$0.20$k_\mathrm{B} T_{\rm c}$, which is comparable to the lowest temperature ($T=$~0.27$T_{\rm c}$).
Therefore, some effects of the magnetic field are expected around the lowest temperature.
As shown in Fig.~\ref{mu-T}, the observed mobility at 0.25~T has the following two features: (1) It is slightly smaller than that in the zero magnetic field at 0.35$T_{\rm c}$~$\lesssim T <$~0.48$T_{\rm c}$. (2) The rate of the increase of the mobility is larger than that in zero magnetic field at temperatures below $\sim$~0.35$T_{\rm c}$, and the mobility is expected to be higher below 0.27$T_{\rm c}$. 
These features are explained as a result of the gap opening of the Majorana surface states and the distortion of the order parameter of bulk $^3$He-B as discussed below.

\section{Theory: mobility of electron bubble in magnetic field}

The mobility of the electron bubbles is determined by the momentum transfer from the $^3$He quasiparticles to the electron bubble.
The equation of motion for the momentum of the electron bubble moving parallel to the surface is given by $d{\bm P}/dt=-\eta_{\parallel }{\bm v}_{\parallel }$ up to the first order of the velocity of the electron bubble ${\bm v}_{\parallel}$~\cite{bromley:1981,shevtsov:2016,Tsutsumi2017}.
The Stokes drag coefficient $\eta_{\parallel}$ is obtained by
\begin{align}
\eta_{\parallel }=\frac{\pi\hbar }{2}\sum_{{\bm k},{\bm k}'}({\bm k}_{\parallel }'-{\bm k}_{\parallel })^2\left(-\frac{\partial f_{\bm k}}{\partial E_{\bm k}}\right)\delta(E_{{\bm k}'}-E_{\bm k})
\sum_{\sigma,\sigma'}|t({\bm k},\sigma\to{\bm k}',\sigma')|^2,
\label{eq:eta1}
\end{align}
where $f_{\bm k}\equiv f(E_{\bm k})=[1+\exp(E_{\bm k}/k_{\rm B}T)]^{-1}$ is the Fermi distribution for the quasiparticle excitation energy $E_{\bm k}$.
For the surface bound state in $^3$He-B, the spinors for the Majorana fermions are described by $\Psi_+(\phi)=(1,-i e^{i\phi},-e^{i\phi},-i)^{\rm T}/\sqrt{2}$ and $\Psi_-(\phi)=(e^{-i\phi},i,1,-ie^{i\phi})^{\rm T}/\sqrt{2}$ for the quasiparticle energy $\pm\Delta_{\rm B}\sin\theta$~\cite{Chung2009,nagato:2009}, where $\phi$ and $\theta$ are the azimuthal angle and the polar angle from the normal axis to the free surface, respectively.
When the magnetic field $H$ is applied perpendicular to the surface, the surface bound state is given by $|\Psi_{\bm k}\rangle=\Psi_H|{\bm k}\rangle$ with $\Psi_H\equiv a\Psi_+(\phi)+b\Psi_-(\phi)$~\cite{nagato:2009}.
The coefficients are $a=e^{-i\phi}(\hbar\gamma H/2)/\sqrt{(\Delta_{\rm B}\sin\theta-E_{\bm k})^2+(\hbar\gamma H/2)^2}$ and $b=(\Delta_{\rm B}\sin\theta-E_{\bm k})/\sqrt{(\Delta_{\rm B}\sin\theta-E_{\bm k})^2+(\hbar\gamma H/2)^2}$ for the gapped energy $E_{\bm k}=\pm\sqrt{(\Delta_{\rm B}\sin\theta)^2+(\hbar\gamma H/2)^2}$.
The gapped surface bound state gives the squared $T$-matrix element as $\sum_{\sigma,\sigma'}|t({\bm k},\sigma\to{\bm k}',\sigma')|^2=\left|\langle\Psi_{{\bm k}'}|T_{\rm S}|\Psi_{\bm k}\rangle\right|^2$.
The wave number of the surface bound state is fixed on the Fermi wave number, ${\bm k}=k_{\rm F}\hat{\bm k}$.
The $T$-matrix element $\langle{\bm k}'|T_{\rm S}|{\bm k}\rangle\equiv T_{\rm S}(\hat{\bm k}',\hat{\bm k},E,z)$ is given by the following equation based on the Lippman-Schwinger equation~\cite{salomaa:1980,shevtsov:2016,Tsutsumi2017}:
\begin{multline}
T_{\rm S}(\hat{\bm k}',\hat{\bm k},E,z)\\
=T_{\rm N}(\hat{\bm k}',\hat{\bm k})
+N_{\rm F}\int\frac{d\Omega_{{\bm k}''}}{4\pi }T_{\rm N}(\hat{\bm k}',\hat{\bm k}'')
\left[g_{\rm S}(\hat{\bm k}'',E,z)-g_{\rm N}\right]T_{\rm S}(\hat{\bm k}'',\hat{\bm k},E,z),
\label{eq:LS}
\end{multline}
where $N_{\rm F}$ is the density of states in the normal state per spin.
Since the size of the electron bubble is much smaller than the coherence length, the $T$-matrix depends on the quasiclassical Green's function $g_{\rm S}(\hat{\bm k},E,z)$ only at the position of the electron bubble $z$~\cite{thuneberg:1981}.
The quasiclassical Green's function, which includes the contributions from poles of the retarded Green's function near the Fermi surface, is defined by $g(\hat{\bm k},E,z)\equiv\oint d\xi_{\bm k}G({\bm k},E,z)$~\cite{kopnin:book}.
The retarded Green's function consists of the wave functions $\Psi_{\nu}({\bm r})$ with energy $E_{\nu}$ as $G({\bm r}_1,{\bm r}_2,E)=\sum_{\nu}\frac{\Psi_{\nu}({\bm r}_1)\Psi_{\nu}({\bm r}_2)^{\dagger}}{E-E_{\nu}+i0^+}$, where $\nu$ is the index of an eigenstate.
The part of the Green's function from the surface bound state is given by the wave function $\Psi_{\bm k}({\bm r})=\Psi_H\langle{\bm r}|{\bm k}\rangle$ with energy $E_{\bm k}$. 
The Green's function for $E_{\nu}>\Delta_{\rm B}$ is regarded as that in the bulk under magnetic field.
Note that the quasiclassical Green's function in the normal state is given by $g_{\rm N}=-i\pi\tau_0$.
Note also that the distortion of the order parameter of the bulk superfluid $^3$He-B due to the magnetic field was neglected.

\begin{figure}[th]
	\begin{center}
		\includegraphics[width=\linewidth,keepaspectratio]{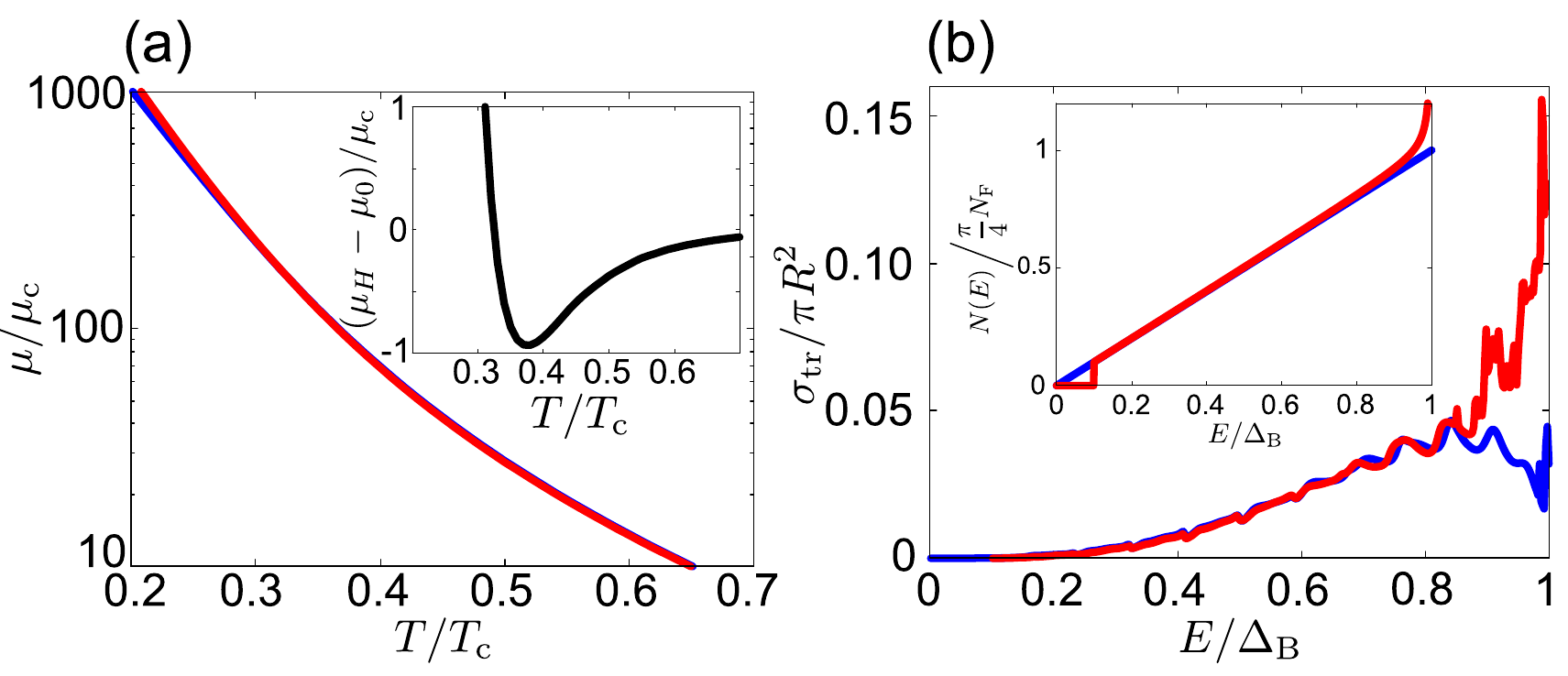}
	\end{center}
	\caption{(a) Calculated mobility of the electron bubbles at $z=40$ nm without magnetic field (blue line, $\mu_0$) and under magnetic field $\hbar\gamma H/2=0.2k_{\rm B}T_{\rm c}$ (i.e., $H=$~0.25~T) (red line, $\mu_H$).
		Inset: Difference of the mobility, $(\mu_H-\mu_0)/\mu_{\rm c}$.
		(b) Transport cross section at $z=0.5\xi$ without magnetic field (blue line) and under magnetic field $\hbar\gamma H/2=0.1\Delta_{\rm B}$ (red line). Here the coherence length is given by $\xi = \xi_0 [\Delta_B(T=0) / \Delta_B(T)]$ with the coherence length at $T=0$, $\xi_0 \approx 137$ nm. Inset: Density of states $N(E)$ at the free surface without magnetic field (blue line) and under magnetic field $\hbar \gamma H/2 = 0.1 \Delta_{\rm B}$ (red line) calculated at zero depth.
	}
	\label{fig:th}
\end{figure}

The $T$-matrix in the normal state is given by $T_{\rm N}(\hat{\bm k}',\hat{\bm k})={\rm diag}[t_{\rm N}(\hat{\bm k}',\hat{\bm k})\sigma_0,-t_{\rm N}(-\hat{\bm k}',-\hat{\bm k})^*\sigma_0]$ with
\begin{align}
t_{\rm N}(\hat{\bm k}',\hat{\bm k})=-\frac{1}{\pi N_{\rm F}}\sum_{l=0}^{\infty }(2l+1)e^{i\delta_l}\sin\delta_lP_l(\hat{\bm k}\cdot\hat{\bm k}'),
\end{align}
where $P_l$ is the Legendre polynomial and $\delta_l$ is the phase shift depending on the potential of the electron bubble.
Here, $\tau_0$ ($\sigma_0$) is the unit matrix in the Nambu- (spin-)space.
By solving Eq.~\eqref{eq:LS}, we derive the squared $T$-matrix element in Eq.~\eqref{eq:eta1} and obtain the mobility $\mu=e/(\eta_{\parallel}+\eta_{\rm B})$ in the same manner as Ref.~\cite{Tsutsumi2017}, where $\eta_{\rm B}$ is the drag coefficient in the bulk $^3$He-B.
The potential of an electron bubble is well modeled by the hard sphere potential~\cite{shevtsov:2016b}.
For the hard sphere potential with radius $R$, the phase shift $\delta_l$ is given by $\tan\delta_l=j_l(k_{\rm F}R)/n_l(k_{\rm F}R)$, where $j_l$ and $n_l$ are the spherical Bessel and Neumann functions, respectively.
The radius $R$ of the hard sphere potential is fixed by the observed mobility at the superfluid transition temperature $\mu_{\rm c}=1.7\times 10^{-6}\ {\rm m^2/Vs}$ as $R=11.17k_{\rm F}^{-1}$~\cite{shevtsov:2016b}.

The mobility of the electron bubble at $z=40$~nm is shown in Fig.~\ref{fig:th}(a) without magnetic field (blue line) and under magnetic field $\hbar\gamma H/2=0.2k_{\rm B}T_{\rm c}$ (i.e., $H=$~0.25~T) (red line).
The mobility under the magnetic field, $\mu_H$, is larger than that without magnetic field, $\mu_0$, at low temperatures because the surface bound state has a gap $\hbar\gamma H/2$.
As shown in the inset of Fig.~\ref{fig:th}a, the values of the mobility are reversed around $T=0.32T_{\rm c}$, which is around the lowest temperature in the experiment (Fig.~\ref{mu-T}).

The suppression of the mobility under the magnetic field above $T=0.32T_{\rm c}$ is owing to the enhancement of the transport cross section for high energy quasiparticles as shown in Fig.~\ref{fig:th}(b).
The transport cross section $\sigma_{\rm tr}(E,z)$ is related to the drag force as~\cite{Tsutsumi2017}
\begin{align}
\eta_{\parallel}=n_3p_{\rm F}\int_{-\Delta_{\rm B}}^{\Delta_{\rm B}}dE\left(-\frac{\partial f}{\partial E}\right)\sigma_{\rm tr}(E,z)\left[\frac{\pi}{4}{\rm sech}^2\left(\frac{z}{2\xi}\right)\right]^2,
\end{align}
where $n_3=k_{\rm F}^3/3\pi^2$ is the $^3$He density, $p_{\rm F}$ is the Fermi momentum, and the spatial dependence $\frac{\pi}{4}{\rm sech}^2\left(\frac{z}{2\xi}\right)$ with the coherence length $\xi\equiv\hbar v_{\rm F}/2\Delta_{\rm B}$, ($v_{\rm F}$ is the Fermi velocity), originates from the density of states for the surface bound state~\cite{tsutsumi:2012}.
The surface bound states under the magnetic field have the relatively large density of states at $0.8\Delta_{\rm B} \lesssim |E|\lesssim\Delta_{\rm B}$ because the states in this energy region are moved from $|E|<\hbar\gamma H/2$ (see Fig.~\ref{fig:th}b inset).
This leads to the enhancement of $\sigma_{\rm tr}$ at $0.8\Delta_{\rm B} \lesssim |E|\lesssim\Delta_{\rm B}$ (Fig.~\ref{fig:th}b),
This enhancement is relatively large because the surface bound states around $\Delta_{\rm B}$ have a large momentum parallel to the surface.
Thus, the mobility under the magnetic field is suppressed at $T>0.32T_{\rm c}$ (Fig.~\ref{fig:th}a).

The calculated mobility under magnetic field approaches the mobility without magnetic field at high temperatures because the distortion of the order parameter of bulk $^3$He-B is disregarded in this calculation.
The gap perpendicular to the surface, $\Delta_{\perp}$, is suppressed as $\Delta_{\perp}\approx 0.9\Delta_{\rm B}$ by magnetic field around 0.25 T at $T\lesssim0.3T_{\rm c}$ and gradually decreases as raising temperature~\cite{ashida:1985}.
The quasiparticles with energy above $\Delta_{\perp}$ cause the drag force of the electron bubbles similarly to the quasiparticles around the gap nodes in $^3$He-A~\cite{salomaa:1980,shevtsov:2016}.

\section{Discussion}

As seen in Fig.~\ref{mu-T}, even at our lowest temperature (0.27$T_{\rm c}$), we do not observe a large difference in mobility from that at zero magnetic field, although the Zeeman gap $\hbar \gamma H /2(=$~0.20$k_\mathrm{B} T_{\rm c}=$~0.12$\Delta_\mathrm{B}$ at 0.25~T) is comparable to the temperature.
This is because, at $E \lesssim \hbar \gamma H /2$($=$~0.12$\Delta_\mathrm{B}$), the transport cross section for the surface Andreev bound states at zero magnetic field is small (see Fig.~\ref{fig:th}b).
Therefore, even if the Zeeman gap opens, the change in the mobility is rather small. 
However, the absence of the cross section within the Zeeman gap leads to a steeper increase of the mobility in the magnetic field, as observed below $\sim$~0.3$T_{\rm c}$ in the experiment (Fig.~\ref{mu-T}) and in the theory (Fig.~\ref{fig:th}a).
This suggests that the difference in mobility between 0~T and 0.25~T is expected to be larger at further lower temperatures, and thus the Zeeman gap can be observed at such temperature range.

The experimental mobility at 0.25~T is smaller than that in the zero magnetic field at 0.35$T_{\rm c}$~$\lesssim T <$~0.48$T_{\rm c}$ (Fig.~\ref{mu-T}).
There are two reasons for this.
One is that, in the magnetic field, the density of states in the Zeeman gap moves to around the superfluid gap $\Delta_\mathrm{B}$.
This leads to the enhancement of the transport cross section at $0.8 \lesssim E/\Delta_\mathrm{B} \lesssim 1$ in Fig.~\ref{fig:th}b as discussed in Sect.~4, which gives rise to the smaller mobility in the magnetic field as seen at $\sim$~0.38$T_{\rm c}$ in the inset of Fig.~\ref{fig:th}a.
The second reason is that the bulk superfluid gap $\Delta_\mathrm{B}$ distorts elliptically in the magnetic field~\cite{Vollhardt90}.
In our experimental configuration, the gap becomes smaller in the direction perpendicular to the surface, while it becomes larger in the direction parallel to the surface. 
This distortion, which is not included in the theoretical calculation in Sect.~4, causes more excitations of quasiparticles in the direction perpendicular to the surface, resulting in the suppression of the mobility from that in the zero magnetic field, as observed at 0.35$T_{\rm c}$~$\lesssim T <$~0.48$T_{\rm c}$ in Fig.~\ref{mu-T}.
At lower temperatures, the density of the bulk quasiparticles decreases rapidly and the mobility is more dominated by the scattering with the surface Andreev bound states.

\section{Conclusions}
We study the mobility of an electron bubble trapped under a free surface of $^3$He-B in a magnetic field of 0.25~T.
Our experiment and theory show that the mobility at 0.25~T increases steeper than that in zero magnetic field with decreasing temperature at $T \lesssim $~0.3$T_{\rm c}$, reflecting the Zeeman gap of the surface Andreev bound states.
At higher temperatures, the mobility is found to be slightly smaller than that in zero magnetic field.  
This is explained by the change in the density of states within the bulk superfluid gap and the distortion of the bulk superfluid gap by the magnetic field.

Our results indicate that the mobility of the electron bubble is not so sensitive to the Zeeman gap of the surface Andreev bound states at the investigated temperature range.
This is due to the small cross section of the electron bubble for surface Andreev bound states at low energies.
Investigating the mobility at further lower temperatures is one of the directions to clearly observe the Zeeman gap.
Another root is to find a method that can sensitively detect properties of the low-energy states by using electron bubbles.

\backmatter

\bmhead{Acknowledgments}
We thank J.A. Sauls, Y. Nagato, S. Higashitani, T. Mizushima, and S.-B Chung for illuminating discussions. KK is grateful to Prof. Ben-Li Young and Prof. Wen-Bin Jian for hospitality.

\bmhead{Author Contributions}
H.I. and K.K. carried out the experiments and Y.T. performed the theoretical calculations. All authors contributed equally to the preparation of the manuscript.

\bmhead{Funding}
This work is supported by JSPS KAKENHI Grant No. 21K03456 and National Science and Technology Council, Taiwan (Grant No. NSTC 113-2112-M-A49-040-).

\bmhead{Data Availability}
The data sets generated during the current study are available from the corresponding authors on reasonable request.

\section*{Declarations}
K.K. is a member of the Editorial Board of J. Low Temp. Phys. Otherwise, all authors have no financial or proprietary interests in any material discussed in this article. The data acquisition was made at the Low Temperature Physics Laboratory, RIKEN.

\bibliography{3He-B_high_field.bib}

\end{document}